\def\diff#1#2{\frac{\partial #1}{\partial #2}}

\documentclass[12pt,oneside]{amsart}

\usepackage{amsmath}               
\usepackage{times}

\numberwithin{equation}{section}

\title[Anisotropic generalizations of de Sitter spacetime]%
{Anisotropic generalizations of de Sitter spacetime}

\author[R.\ Giamb\`o]{Roberto Giamb\`o}
\address{Dipartimento di Matematica e Informatica,
Universit\`a di Camerino, Italy}
\email{roberto.giambo@unicam.it}
\urladdr{http://www2.unicam.it/\~{}giambo}

\begin{document}

\begin{abstract}
It is known that de Sitter spacetime can be seen as the solution
of field equation for completely isotropic matter. In the present
paper a new class of exact solutions in
spherical symmetry is found and discussed,
such that the energy--momentum tensor has two 2--dimensional
distinct isotropic subspaces.

{\medskip\noindent PACS numbers 04.20.Jb, 04.70.Bw}
\end{abstract}

\maketitle

\section{Introduction}\label{sec:intro}
As is well known, the energy-momentum tensor describing ordinary
matter is of the Pleba\'nski type $[T-S_1-S_2-S_3]_{(1111)}$, i.e. it admits
one timelike and three spacelike eigenvectors ($u^\mu$ and
$(m^a)^\mu$ with $a=1,2,3$, say). The eigenvalue corresponding to
the timelike vector field $u$ is (minus) the energy density of the
matter $\epsilon$, while the three spacelike eigenvectors are the
principal stresses $\lambda_a$ ($a=1,2,3$) (obviously, the perfect
fluid is space-isotropic, so that for a fluid the spacelike
eigenvalues coincide). Using as tetrad basis that defined by the
eigenvectors, the energy momentum tensor can be written in the
form
\begin{equation}\label{eq:T}
T^\mu_\nu=\epsilon u^\mu u_\nu +\sum_a \lambda_a (m^a)^\mu
(m^a)_\nu
\end{equation}

From now on we work in spherical symmetry (see \cite{Kra}), where only two
spacelike eigenvalues can be distinct, so that the canonical form
is
\begin{equation}\label{eq:T--ss}
T^\mu_\nu=\epsilon u^\mu u_\nu +p_r m^\mu m_\nu +p_t
\Delta^\mu_\nu\ ,
\end{equation}
where $m^\mu$ denotes the unit radial vector and $\Delta$ denotes
the projector onto the two-dimensional spacelike subspace
orthogonal to $m$. The radial and tangential stresses are
respectively denoted by $p_r$ and $p_t$.

The de Sitter spacetime can be viewed as a vacuum solution with
nonzero "lam\-bda" term but also, as is well known, as the spacetime
originated by a lambda term "source", i.e. as the solution of the
Einstein field equation in matter having
$T^\mu_\nu=-\epsilon_0 \delta^\mu_\nu$ where
$\epsilon_0$ is now constant due to the field equations for
matter, which obviously imply $\partial_\nu \epsilon_0=0$. In the
formula \eqref{eq:T--ss},
one thus has $\epsilon=-p_t=-p_r=\epsilon_0$ so that
the energy tensor is completely degenerate.

One may now ask if there are ways, in which it is possible to
weaken the hypothesis of complete degeneracy in a "minimal" way,
i.e. retaining two degenerate subspaces, thereby obtaining
"anisotropic generalizations" of de Sitter spacetime. It is obvious, that
perfect fluids cannot achieve this goal. However, if the stress is
anisotropic then one can search for solutions described by
\begin{equation}\label{eq:T--now}
T^\mu_\nu= -A \Gamma^\mu_\nu +p_t \Delta^\mu_\nu
\end{equation}
where $A$ is some function and $\Gamma$ is the unit tensor living
in the two-dimensional subspace spanned by the timelike and the
radial spacelike eigenvectors (i.e. $\Gamma^\mu_\nu =-u^\mu u_\nu
+m_\mu m^\nu$).
Interestingly enough, the equations of
motion for the matter now shows that $\epsilon$ is a function
depending on $R$ only, where $R$ is the comoving area radius.

In the present paper this class of solutions is derived,
together with minimal physical requirements that have to be
imposed. It also includes models used to build up
regular black hole interiors \cite{Dym, EH, Mag2, Barca}, where a
regular solution is used to replace a singular core (e.g. Schwarzschild).
It may be worthwhile noticing that here the starting point
is given by a condition on the constitutive equation \eqref{eq:state},
(see \eqref{eq:condcost} below)
recovering a solution which in principle is neither static nor regular.

\section{The solution}\label{sec:sol}

Consider a spherically symmetric object, whose general line
element in comoving coordinates may be written as
\begin{equation}\label{eq:ds}
\text ds^2=-e^{2\nu}\text dt^2 +e^{2\lambda}\text dr^2 +R^2 (\text d\theta^2
+\sin^2\theta\, \text d\phi^2)
\end{equation}
(where $\nu, \lambda$ and $R$ are function of $r$ and $t$). With
the energy--momentum tensor describing the matter given by
\eqref{eq:T--ss}, Einstein field equations for this model reads
\begin{subequations}
\begin{align}
& m'=4\pi\epsilon\, R^2\, R',\qquad\dot m=-4\pi\,p_r\, R^2\,\dot R,\label{eq:E12}\\
&\dot R'=\dot\lambda R'+\nu'\dot R,\label{eq:E3}\\
&p_r'=-(\epsilon+p_r)\,\nu'-{2 \frac{R'}R}(p_r-p_t),\label{eq:E4}
\end{align}
\end{subequations}
where a prime and a dot denote partial derivative with respect to
$r$ and $t$ respectively, and $m$ is the {\sl Misner--Sharp} mass,
defined as
\begin{equation}\label{eq:M-S}
m(r,t)={\frac R2}\left[1-(R')^2\,e^{-2\lambda}+\dot R^2
e^{-2\nu}\right].
\end{equation}
The equation of state for a general material in spherical symmetry
can be given in terms of a {\sl state} function (see e.g.
\cite{Mag})
\begin{equation}\label{eq:state}
\epsilon=\epsilon(r,R,\eta),
\end{equation}
where $\eta=e^{-2\lambda}$, in such a way that the stresses, which
are in general anisotropic, are given by the following relations:
\begin{equation}\label{eq:stress}
p_r=-\epsilon+2\eta \diff{\epsilon(r,R,\eta)}\eta, \qquad p_t=-\epsilon
-\frac R2\,\diff{\epsilon(r,R,\eta)}R.
\end{equation}
It follows,
that since, from \eqref{eq:T--now},
the solutions we are searching for are uniquely
characterized by the condition that
\begin{equation}\label{eq:cond}
\epsilon+p_r=0,
\end{equation}
then $\epsilon$ has to be independent from $\eta$:
\begin{equation}\label{eq:condcost}
\diff\epsilon\eta=0.
\end{equation}
Substituting \eqref{eq:stress} in \eqref{eq:E4} and using
$\epsilon'(r,t)=\diff{\epsilon(r,R)}r+
\diff{\epsilon(r,R)}R R'$ coming from \eqref{eq:condcost}, we get
also $\diff\epsilon r=0$, that is $\epsilon$ given by \eqref{eq:state} must be a
function of $R$ only.
We are thus left with the following
energy-momentum tensor:
\begin{equation}\label{eq:T--aniso}
T^\mu_\nu= -\epsilon\,\Gamma^\mu_\nu -(\epsilon +\frac R2\, \frac{\text d\epsilon}{\text dR} )
\Delta^\mu_\nu,
\end{equation}
and since equations
\eqref{eq:E12} now read
\[
m'=4\pi\epsilon(R)\, R^2\, R',\qquad \dot
m=4\pi\epsilon(R)\,R^2\dot R,
\]
it must be
\begin{equation}\label{eq:mass}
m(R)=4\pi\int_0^R\epsilon(\sigma)\sigma^2\,\text d\sigma+m_0.
\end{equation}
In order for this solution to satisfy minimal requirements of
acceptability, the {\sl weak energy condition (w.e.c.)}
is imposed, that reads
\begin{equation}\label{eq:wec}
\epsilon\ge 0,\quad\epsilon+p_r\ge 0,\quad\epsilon+p_t\ge 0
\end{equation}
(a basic reference for a discussion of energy conditions is
\cite{HE}).
In this case \eqref{eq:stress} implies that w.e.c. is satisfied if
$\epsilon(R)$ is a non negative and not increasing function of
$R$.

We are left with the system \eqref{eq:E3}, \eqref{eq:M-S}, and
\eqref{eq:mass} in the unknown $R,\,\lambda,\,\nu,\,\epsilon$ and
$m$.
In principle \eqref{eq:E3} should be integrated in order to obtain
exact solutions, but
recalling, from \eqref{eq:M-S} and
\eqref{eq:mass}, that $(R'e^{-\lambda})^2-(\dot Re^{-\nu})^2$ is a
function of $R$ only, we will
limit ourselves to the case when the two addenda are
separately function of $R$ only:
\begin{equation}\label{eq:ab}
R'e^{-\lambda}\equiv a(R),\qquad \dot Re^{-\nu}\equiv b(R),
\end{equation}
where $a$ and $b$ are two {\sl prescribed} functions of $R$.

Using condition \eqref{eq:E3} we find, up to time
reparameterization, the following expressions:
\begin{align*}
&\lambda=\log(f(r)b(R)),\quad\nu=\log a(R),\\
&m(R)=\frac R2[1-a^2(R)+b^2(R)],\quad\epsilon(R)=\frac 1{4\pi R^2}
\frac{\text dm(R)}{\text dR},
\end{align*}
and the two (compatible) equations for $R(r,t)$:
\begin{equation}\label{eq:R}
\dot R=a(R) b(R),\qquad R'=f(r) a(R) b(R).
\end{equation}
Here the function $f(r)$ arises as an integration term, and since
the curve $R(r,0)$ is the initial data that will be conveniently
taken equal to $r$, it is $R'(r,0)=1$ and then
$f(r)=(a(r)b(r))^{-1}$.

With the position
\[
v(R):=a(R)b(R),
\]
equations \eqref{eq:R} reads
\begin{equation}\label{eq:R1}
\dot R=v(R),\qquad R'=\frac{v(R)}{v(r)},
\end{equation}
that integrate to give
\begin{equation}\label{eq:R-now}
R(r,t)=V^{-1}(t+V(r)),
\end{equation}
where $V(\sigma)$ is a primitive of $\frac 1{v(\sigma)}$
(note that $V$ is invertible). The line element then takes the
form
\begin{equation}\label{eq:sol}
\text ds^2=-a^2(R)\text dt^2+\frac{v^2(R)}{v^2(r)\,a^2(R)}\text dr^2+R^2\text
d\Omega^2,
\end{equation}
with $R=R(r,t)$ given by \eqref{eq:R-now}.

\section{Physical interpretation and remarks}
Defining the function
\begin{equation}\label{eq:chi}
\chi (R)=1-\frac{2m(R)}{R}
=a^2(R)-b^2(R)=a^2(R)-\frac{v^2(R)}{a^2(R)},
\end{equation}
it can be seen that two
suitable coordinate changes map the regions $\{\chi>0\}$ and $\{\chi<0\}$
respectively into a {\sl variation of mass} of Schwarzschild solution.
The sign of $\chi$ defines the character of $R$ viewed as a
coordinate: if $\chi$ is positive, $R$ may be regarded as a
''length", if $\chi$ is negative $R$ can be seen as a ''time"
instead.

It can be noticed that these
solutions can arise from Kerr-Schild geometry, since another
suitable coordinate change may be applied,
to recover the form
\begin{equation}\label{eq:KS}
\text ds^2=(-\text d\overline t^2+\text d\overline r^2+\overline
r^2\text d\Omega^2)+(1-\chi(\overline r))(\text d\overline t+\text
d\overline r)^2,
\end{equation}
that shows, by the way, that $\chi=0$ is a
removable singularity.
The family outlined here in fact contains Minkowski, Schwarzschild and de
Sitter spacetimes as particular cases, corresponding to choosing the function $m(R)$
respectively equal to 0, to $m_0$ (constant), and to $\frac 43\epsilon_0 \pi R^3$.

As sketched before, the line element of these metrics are formally
analogue to Schwarzschild exterior and black hole (coinciding with
them if $m(R)$ is constant), so it is not surprising that $\chi=0$
is not a true singularity. Computation of Kretschmann scalar $K=R_{abcd}R^{abcd}$
yields
\begin{equation}\label{eq:Kre}
K=4\frac{(\chi-1)^2}{R^4}+4\frac{(\chi')^2}{R^2}+(\chi'')^2,
\end{equation}
and this suggest that the only singularity can occur at $R=0$.
Let us assume $m$ (and therefore $\chi$) analytic in a right neighborhood of 0,
and let $\alpha\ge 0$ such that $m(R)\cong R^\alpha$.
It is easily seen that metric regularity at $R=0$ occurs if and only if
$\alpha\ge 3$. On the other side, since
w.e.c. \eqref{eq:wec} in terms of $m$ reads
\begin{equation}\label{eq:wec2}
m'(R)\ge 0,\qquad m''(R)-\frac 2R m'(R)\le 0,
\end{equation}
in order for \eqref{eq:wec2} to be satisfied near $R=0$ it must be
$\alpha\le 3$. Therefore, the only case for a physically acceptable
metric not to be singular at $R=0$ is
when it behaves asymptotically as de Sitter
spacetime as $R$ approaches 0 (that is, $m(R)\cong R^3$). In all
other cases (allowed by w.e.c.) $R=0$ is a true singularity.

The solutions here found can be matched to Schwarzschild spacetime
if and only if $\epsilon$ vanishes on a $R=\text{const.}$ surface,
that is there exists $R_b>0$ such that $\frac{\text dm}{\text
dR}(R_b)=0$. This can be performed either in the static region (i.e.
where $\chi>0$), or in the non static one. In the first case, we
obtain a globally static object.

If, instead, we allow $\chi$ to vanish for some $R_0>0$
(Cauchy horizon), the solution may enter the non static region,
where the matching can be performed obtaining physically
valid black hole interior models.
This is not in contradiction
with the result proved by Baumgarte and Rendall in \cite{BR},
stating that the inequality $\chi(R)>0$ remains true until radial
pressure vanishes. Indeed, the hypothesis $\epsilon+p_r>0$
is crucial for the argument used in \cite{BR}, whereas in our case
the w.e.c. is satisfied at its borderline, that is
$\epsilon+p_r\equiv 0$.
As pointed out before,
these solutions are regular only if they have a
de Sitter--like behavior as $R\to 0^+$, other regular choices
being forbidden by w.e.c..
The junction can even be done between a true de Sitter core and a
Schwarzschild spacetime, but this time an intermediate thick shell
is needed in the middle, to ensure continuity of radial pressure.
An example is given by
\begin{subequations}
\begin{equation}
m(R)=
\begin{cases}
\frac43\pi\epsilon_0 R^3,\quad &R\in[0, kM[\\
\frac 43\pi\epsilon_0 (R-2kM)^3+2M,\quad &R\in[kM,2kM[\\
2M,\quad &R\in[2kM,+\infty)
\end{cases}
\end{equation}
where $k$ is a constant such that,
to ensure continuity, the Schwarzschild mass $M$
equals twice the mass of de Sitter $\frac 43\pi\epsilon_0 (kM)^3$.
This choice yields the following energy density function:
\begin{equation}\label{eq:eps-ex1}
\epsilon(R)=
\begin{cases}
\epsilon_0,\quad &R\in[0, kM[\\
\epsilon_0 \left(1-\frac{2kM}{R}\right)^2,\quad &R\in[kM,2kM[\\
0,\quad &R\in[2kM,+\infty).
\end{cases}
\end{equation}
\end{subequations}
The limitation $k\le 1$ gives a relation between the parameters of de
Sitter and Schwarz\-schild solutions (i.e. $\epsilon_0M^2\le \frac 3{8\pi}$)
to be satisfied in order to
obtain a regular black hole interior model with a single Cauchy horizon.

Regular Schwarzschild black holes
have been studied by several authors \cite{Dym, EH, Mag2, Barca},
and explicit solutions have been build up, with a number
of Cauchy horizons even greater than one before energy
density vanishes (see \cite{Barca} for an example).

Of course, matching with Schwarzschild spacetime may be also
performed if $m(R)$ $\cong R^\alpha$ with $\alpha<3$, this operation
resulting in replacing Schwarzschild black hole with a non flat
core still possessing a singularity at $R=0$. As an example, if
$m(R)$ is a second order polynomial in $R$,
\begin{equation}\label{eq:ex1}
m(R)=-kR^2+4kMR,\qquad R\in[0,2M],
\end{equation}
the junction is done at $R=2M$, and the
condition $k\ge\frac1{4M}$ ensures that this matching is made in
the non static region. Therefore, in this case there are no Cauchy horizons.

We also notice that the above constructions are made supposing
continuity of both the metric and radial pressure $p_r$.
For instance, if the core is a Schwarzschild solution, we cannot
perform a matching with one of our solution at $R_0>0$,
since continuity of $p_r$ is lost at $R_0$. Anyway, if
discontinuities of $p_r$ at junctions are allowed,
other spacetimes can be built.
Indeed, matching between two variations of mass of Schwarzschild
spacetime --
generated by an "inside--shell" mass $m(R)$ and an
"outside--shell" mass $M(R)$ -- can be performed at
$R_0$ such that, for instance, both $\chi_m(R_0)$ and $\chi_M(R_0)$
are positive.
In this case it must be checked that the surface stress--energy
tensor (i.e. related to the surface $\mathcal S=\{R=R_0\}$)
obeys weak energy condition. Following the method in \cite{MTW}
the surface energy density and pressure are found to be
\begin{subequations}
\begin{align}
&\epsilon_{\mathcal S}=\frac{1}{4\pi
R_0}\left(\sqrt{\chi_m(R_0)}-\sqrt{\chi_M(R_0)}\,\right),\\
&p_{\mathcal S}=\frac{1}{8\pi
R_0^2}\left[\frac{(1-M'(R_0))R_0-M(R_0)}{\sqrt{\chi_M(R_0)}}-
\frac{(1-m'(R_0))R_0-m(R_0)}{\sqrt{\chi_m(R_0)}}\right],
\end{align}
\end{subequations}
and for the solution to be physically acceptable it must be
$\epsilon_{\mathcal S}\ge 0$ (that is, $M(R_0)\ge m(R_0)$) and
$\epsilon_{\mathcal S}+p_{\mathcal S}\ge 0$.
A particular
case of a Schwarzschild region surrounded by a massive shell
is studied in \cite{Fra}, where the region outside shell
can also be viewed as a Schwarz\-schild solution,
under a suitable coordinate change.

\end{document}